\documentclass[11pt]{article}

\usepackage[dvips]{graphicx}
\usepackage{epsf}  
\usepackage{epsfig}  

\newcommand{\BABARPubYear}    {04}

\newcommand{\BABARConfNumber} {40}
\newcommand{\SLACPubNumber} {10600}

\input pubboard/babarsym

\newcommand{\psfile}[3][]{ 
  \begin{center}
    \setlength{\epsfxsize}{#3\linewidth}\leavevmode
    \def\noOpt{}\def\testit{#1}\ifx\testit\noOpt%
      \epsfbox{#2}%
    \else%
      \epsfbox[#1]{#2}%
    \fi
  \end{center}
}

\providecommand{\xf}{\mbox{${\cal F}$}}

\providecommand{\DE}{\ensuremath{\Delta E}}
\providecommand{\pvec}{{\bf p}}
\providecommand{\half}{\mbox{${1\over2}$}}

\providecommand{\etapKz}{\mbox{$\eta^{\prime} K^0$}}
\providecommand{\etapKzs}{\mbox{$\eta^{\prime} K^0_S$}}

\providecommand{\kzs}{\ensuremath{\KS}}

\providecommand{\EtapEtaPiPi}{\mbox{$\etapr \rightarrow \eta \pi^+ \pi^-$}}
\providecommand{\EtapRhoPi}{\mbox{$\etapr \rightarrow \rho^0  \gamma$}}

\providecommand{\EtaGG}{\mbox{$\eta \rightarrow \gamma  \gamma$}}
\providecommand{\EtapTrePi}{\mbox{$\eta \rightarrow \pi^+ \pi^- \pi^0$}}

\providecommand{\mgg}{\mbox{$m_{\gamma \gamma}$}}
\providecommand{\etagg}{\mbox{$\eta_{\gamma  \gamma}$}}
\providecommand{\mpipi}{\mbox{$m_{\pi \pi}$}}
\providecommand{\mpipipi}{\mbox{$m_{\pi \pi \pi}$}}
\providecommand{\metap}{\mbox{$m_{ \etapr }$}}

\providecommand{\KSZZ}{\mbox{$K^0_S \rightarrow \pi^0 \pi^0$}}

\providecommand{\tcp}{\mbox{$t_{CP}$}}
\providecommand{\ttag}{\mbox{$t_{\rm tag}$}}

\providecommand{\BetapK}{\mbox{$B^+ \rightarrow \eta^{\prime} K^+ $}}
\providecommand{\BetapKz}{\mbox{$B^0 \rightarrow \eta^{\prime} K^0$}}
\providecommand{\BetapKzs}{\mbox{$B^0 \rightarrow \eta^{\prime} K^0_S$}}

\providecommand{\UfourS}{\mbox{$\Upsilon(4S)$}}
\providecommand{\epem}{\mbox{$e^+e^-$}}
\def\BB{\mbox{$B\overline B\ $}}
\def\pep2{PEP-II}

\providecommand\etal{{\it et al.}}

\providecommand{\progtp}    [1]  {{Prog.\ Theor.\ Phys.\ {\bf #1}}}

\long\def\inst#1{\par\nobreak\kern 4pt\nobreak
    {\it #1}\par\vskip 10pt plus 3pt minus 3pt}

\begin{document}
{\pagestyle{empty}


\begin{flushright}
\babar-CONF-\BABARPubYear/\BABARConfNumber \\
SLAC-PUB-\SLACPubNumber \\
August 2004 \\
\end{flushright}

\par\vskip 3cm

\begin{center}
\Large \bf Measurement of Time-Dependent {\boldmath \CP }-Violating Asymmetries in 
{\boldmath \Bz }  Meson Decays to {\boldmath \etapKz }
\end{center}
\bigskip

\begin{center}
\large The \babar\ Collaboration\\
\mbox{ }\\
\today
\end{center}
\bigskip \bigskip

\begin{center}
\large \bf Abstract
\end{center}

We present a  preliminary measurement of \CP -violating parameters $S$ and $C$ 
from fits of the time-dependence of \Bz\ meson decays to \etapKz . The data 
were recorded with the \babar\ detector at \pep2\ and correspond to 
$227  \times 10^6$ \BB\ pairs produced in \epem\ annihilation through the
\UfourS\ resonance. From a maximum likelihood fit we measure the \CP -violation parameters $S = 0.27 \pm 0.14 $ ({\rm stat}) $ \pm\ 0.03 $ ({\rm syst}) and
 $C = -0.21 \pm 0.10 $ ({\rm stat}) $\pm\ 0.03$ ({\rm syst}).

\vfill
\begin{center}

Submitted to the 32$^{\rm nd}$ International Conference on High-Energy Physics, ICHEP 04,\\
16 August---22 August 2004, Beijing, China

\end{center}

\vspace{1.0cm}
\begin{center}
{\em Stanford Linear Accelerator Center, Stanford University, 
Stanford, CA 94309} \\ \vspace{0.1cm}\hrule\vspace{0.1cm}
Work supported in part by Department of Energy contract DE-AC03-76SF00515.
\end{center}

} 
\newpage
\begin{center}
\small

The \babar\ Collaboration,
\bigskip

%
B.~Aubert,
R.~Barate,
D.~Boutigny,
F.~Couderc,
J.-M.~Gaillard,
A.~Hicheur,
Y.~Karyotakis,
J.~P.~Lees,
V.~Tisserand,
A.~Zghiche
\inst{Laboratoire de Physique des Particules, F-74941 Annecy-le-Vieux, France }
% ...

\end{center}\newpage

\section{Introduction}
\label{sec:Introduction}
Measurements of time-dependent \CP\  asymmetries in $B^0$ meson decays through
a Cabibbo-Kobayashi-Maskawa (CKM) favored $b \rightarrow c \bar{c} s$ amplitude
\cite{babar} have provided a crucial test of the CKM mechanism of \CP\ violation
in the Standard Model (SM)
 \cite{SM}.  Such decays to a charmonium state plus a $K^0$ meson are dominated by a single weak phase.
Decays of  $B^0$ mesons to charmless hadronic final states, such as
$\phi K^0$, $\eta^{\prime} K^0$, $K^+ K^- K^0$, $K^0 \pi^0$ and $f_0(980) K^0$,
are expected to be dominated by penguin diagrams. If we neglect CKM-suppressed 
amplitudes,
these decay modes have the same weak phase as the charmonium $K^0$ decays in the SM.  Thus the time-dependent asymmetry measurement for these decays should yield 
an alternative measurement of $\sin 2 \beta$ \cite{lonsoni}.

\begin{figure}[htbp]
\begin{center}
\vspace{3cm}
\includegraphics[bb=85 155 535 605,angle=0,scale=0.8]{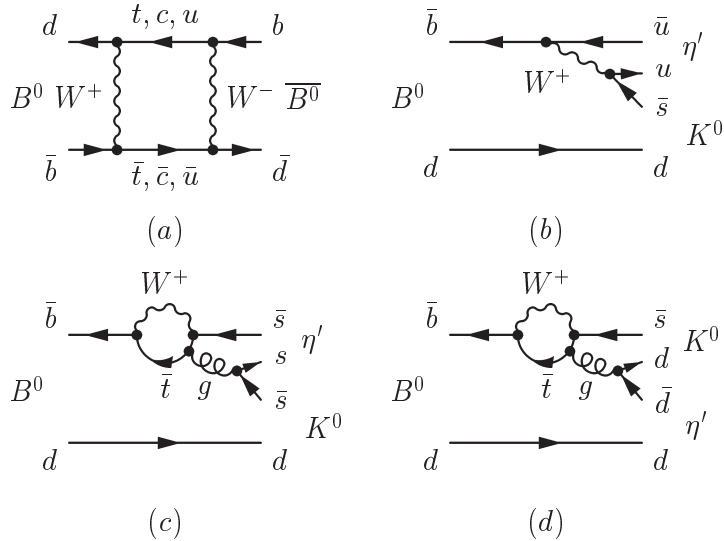}
\vspace{-7cm}
\caption{Feynman diagrams describing (a) $B-\bar{B}$ mixing; the decay
 $B^0 \rightarrow \eta^{\prime} K^0$ via (b) color-suppressed tree, (c,
d) internal gluonic penguin.}
\end{center}
  \label{fig:Feyn}
\end{figure}

The processes shown in  Fig.~\ref{fig:Feyn}(b)-(d) are relevant for the
decay \BetapKz, and there are similar diagrams for \BetapK. 

All of the amplitudes for these processes have CKM suppression, but the tree 
diagram for $B^0$ shown in Fig.~\ref{fig:Feyn}(b) is expected
to be smaller \cite{soni,beneke} since there is additional CKM suppression
and color suppression. For the charged mode the corresponding
tree diagram is external and  not color suppressed.

Additional higher-order amplitudes
carrying  different weak phases would lead to deviations, $\Delta S$, between the measurements of the
time-dependent \CP\ violating parameter in these rare decay modes and in the charmonium $K^0$ 
decays.  Theoretical bounds for these  deviations have been 
calculated with an SU(3) analysis 
\cite {Gross,Gronau}. Such bounds have been improved by recent measurements of
 \Bz\ decays to a pair of neutral charmless light pseudoscalar mesons \cite{Isosca}.
From this and other recent experimental measurements, improved 
model-independent correlated bounds in the ($S$, $C$) plane for the
decay \BetapKz\ have been derived \cite{Gronau2}, with the
conclusion that $\Delta S$ is expected to be less than 0.10 (with a theoretical
uncertainty up to $\sim$30\%  due to the assumptions in the calculation).
Specific model calculations conclude that $\Delta S$ is even smaller;
for instance a recent calculation \cite{BN} finds 
$\Delta S=0.011\pm0.009\pm0.010$, where the first error is due to theoretical
uncertainties for \BetapKz\ and the second is for uncertainties in the 
charmonium-\KS\ system.
A significantly larger value of $\Delta S$ could arise from
phases from non-SM amplitudes \cite{lonsoni}.

The  \CP -violating asymmetry in the decay \BetapKz\  has been  measured previously by the \babar\ ~\cite{Previous}
and Belle ~\cite{BELLE} experiments.  The measurement presented  in this paper is an update of the  
previous \babar\ measurement. In the present analysis the measurements of time-dependent  
\CP\ violating parameters in \BetapK\ are used as a null control sample for
the corresponding measurements in \BetapKz.

\section{The \babar\ Detector and Dataset}
\label{sec:babar}

The results presented in this paper are based on data collected
in 1999--2004 with the \babar\ detector~\cite{babar2}
at the PEP-II asymmetric $e^+e^-$ collider~\cite{pep}
located at the Stanford Linear Accelerator Center.  An integrated
luminosity of 205~fb$^{-1}$, corresponding to about 
227  million \BB\ pairs, was recorded at the $\Upsilon (4S)$ resonance
(``on-resonance'', center-of-mass energy $\sqrt{s}=10.58\ \gev$).

The asymmetric beam configuration in the laboratory frame
provides a boost of $\beta\gamma = 0.56$ to the $\Upsilon(4S)$.
Charged particles are detected and their momenta measured by the
combination of a silicon vertex tracker (SVT), consisting of five layers
of double-sided detectors, and a 40-layer central drift chamber,
both operating in the 1.5 T magnetic field of a solenoid.
The tracking system covers 92\% of the solid angle in the center-of-mass (CM) frame.

Charged-particle identification (PID) is provided by the average
energy loss (\dedx) in the tracking devices  and
by an internally reflecting ring-imaging
Cherenkov detector (DIRC) covering the central region.
A $K/\pi$ separation of better than four standard deviations ($\sigma$)
is achieved for momenta below 3 \gevc , decreasing to 2.5 $\sigma$ at the highest
momenta in the $B$ decay final states.
Photons and electrons are detected by a CsI(Tl) electromagnetic calorimeter (EMC).
The EMC provides good energy and angular resolutions for detection of photons in
 the range from 30 \mev\ to 4 \gev. The energy and angular resolutions are 3\% 
and 4 \mrad, respectively, for a 1 \gev\ photon.

The flux return for the solenoid is composed of multiple layers of iron
and resistive plate chambers for the identification of muons and long-lived
neutral hadrons.

\section{Event Selection and Analysis Method}
Monte Carlo (MC) simulations  of the signal decay modes, \BB\ 
backgrounds, and detector response are used to establish the event selection criteria.

We reconstruct $B$ meson  candidates by combining a \KS\ or a $K^+$   with an \etapr\ meson.
 We select $\KS\to\pi^+\pi^-$ decays by requiring the invariant $\pi^+ \pi^-$ to be within a mass window  of 12 \mev\ 
around the nominal \KS\ mass and  requesting a  flight length $>$ 3 $\sigma$. We  
select $\KS\to\piz\piz$ decays requiring the invariant $\pi^0 \pi^0$ to be within 
a mass window of 30 \mev\ around the nominal \KS\ 
mass and a fit of the two \piz\ mesons to a common decay vertex. We reconstruct 
\etapr\ mesons  through the decays  \EtapRhoPi\ and     \EtapEtaPiPi\   with \EtaGG\  or \EtapTrePi  .
The photon energy $E_{\gamma}$ must be greater than 50 (30)  \mev\ for $\eta$ ($\pi^0$)   candidates,
and greater than 100 \mev\ in  \EtapRhoPi . We make the following 
requirements on the invariant mass (in \mev ): 490 $<$ \mgg $<$ 600  for 
\etagg ,  $120 < \mgg  < 150$  for $\pi^0$  ($100 < \mgg  < 155$ in \KSZZ ), 
$510 < \mpipi\ < 1000$ for $\rho^0$, $520 < \mpipipi\ < 570$  for $\eta_{3\pi}$,
$945<\metap < 970$ for \EtapEtaPiPi, and $930<\metap < 980$ for \EtapRhoPi.

We make several particle identification requirements to ensure the
identity of the signal pions. In  charged $B$ decays  for the $K^+$ track
we require  an associated DIRC Cherenkov angle between
$-5\,\sigma$ and $+2\,\sigma$ from the expected value for a kaon.

A $B$ meson candidate is characterized kinematically by the energy-substituted mass
$\mes \equiv  \sqrt{(\half s + \pvec_0\cdot \pvec_B)^2/E_0^2 - \pvec_B^2}$ and the 
energy difference $\DE \equiv E_B^*-\half\sqrt{s}$, where the subscripts $0$ and
$B$ refer to the initial \UfourS\ and to the $B$ candidate, respectively,
and the asterisk denotes the \UfourS\ frame.
We require $|\DE|\le0.2$ GeV and $5.25\le\mes\le5.29\ \gev$.  

To reject  background in continuum $\epem\ra\qqbar$ events ($q=u,d,s,c$) , we make use
of the angle $\theta_T$ between the thrust axis of the $B$ candidate and
that of
the rest of the tracks and neutral clusters in the event, calculated in
the center-of-mass frame.  The distribution of $\cos{\theta_T}$ is
sharply peaked near $\pm1$ for combinations drawn from jet-like $q\bar q$
pairs and is nearly uniform for the isotropic $B$ meson decays; we require
$|\cos{\theta_T}|<0.9$. The remaining continuum background dominates the 
samples and is modeled from sideband data for the maximum likelihood fits.

We use Monte Carlo simulations of \BzBzb\ and \BpBm\ pair production and decay
to look for \BB\ backgrounds. 
From these studies we find evidence for a small \BB\ background component for 
the channels with 
\EtapRhoPi, and we have added a single \BB\ component to the fit.

From a \BB\ pair we reconstruct a \Bz\  decaying into the  final state $f= \etapKzs $ ($B_{CP}$).
We also reconstruct  the vertex of the other $B$ meson  ($B_{\rm tag}$) and identify its flavor.
The time difference $\deltat \equiv \tcp - \ttag$,
where $\tcp$ and $\ttag$ are the proper decay times of the \CP\ and tagged
$B$ mesons, respectively, is obtained from the measured distance between the $B_{CP}$
and  $B_{\rm tag}$ decay vertices and from the boost ($\beta \gamma =0.56$) of 
the \epem beam system. The distribution of \deltat\ is:
\begin{equation}
F( {\deltat}) = \frac{e^{-\left|\deltat\right|/\tau}}{4\tau} \left\{1 \mp {\Delta \omega} \pm (1 -2 \omega) \left[ S_f\sin(\deltamd
\deltat) - C_f\cos(\deltamd\deltat)\right]\right\},
\label{fplusminus}
\end{equation}
where the upper (lower) sign denotes a decay accompanied by a \Bz\ (\Bzb)
 tag, $\tau$ is the mean \Bz\ lifetime, \deltamd\ is the mixing frequency,
and the mistag parameters $\omega$ and $\Delta \omega$ are the average and difference, 
respectively, of the probabilities that 
a true \Bz\ (\Bzb ) meson is tagged as \Bzb\ (\Bz ). The tagging algorithm, based on six tagging categories, is an improved version of what was used in the  previous \babar\ publication
 \cite{Previous}. Separate neural networks are trained to identify 
primary leptons, kaons, soft pions from $D^*$ decays, and high-momentum 
charged particles from \B\ decays.
Each event is assigned to one of the six mutually exclusive tagging 
categories based on the estimated mistag probability and on the 
source of tagging information.

\section{Maximum Likelihood Fit}
We use an unbinned, multivariate maximum-likelihood fit to extract
signal yields and \CP\-violating parameters. We indicate with $j$ the species of 
event: signal, continuum background, or \BB\ background (\EtapRhoPi).
We use four discriminating variables: 
 \mes , \DE , \deltat , and a Fisher discriminant \xf. The Fisher discriminant
combines  five variables: the angles with respect to the beam axis of the $B$ 
momentum and $B$ thrust axis in the \UfourS\ frame, the zeroth and second 
angular moments of the energy flow excluding the $B$ candidate around the 
$B$ thrust axis, and the tagging category.
For each species $j$ and each tagging  category       
$c$, we define a total probability density function (PDF) for event $i$ as
\begin{equation}
{\cal P}_{j,c}^i \equiv  {\cal P}_j ( \mes^i ) \cdot {\cal  P}_j ( \DE^i )
\cdot { \cal P}_j( \xf^i ) \cdot 
{ \cal  P}_j (\deltat^i , \sigma_{\deltat}^i,c)\,,
\end{equation}
where $\sigma_{\deltat}^i$ is the error on \deltat\ for the event $i$.  
With $n_{j}$ defined to be the number of events of the species $j$
and $f_{j,c}$ the fraction of events of species $j$ for each category $c$,
we write the extended likelihood function for all events belonging to category $c$ as
\begin{equation}
{\cal L}_c= \frac{e^{-N_c}}{ N_c!}
           \prod_i^{N_c} (n_{\rm sig}f_{{\rm sig},c}{\cal P}_{{\rm sig},c}^{i}
                   +n_{q\bar{q}} f_{q\bar{q},c}{\cal P}_{q\bar{q}}^{i}
                   +n_{B\bar{B}}f_{{\rm sig},c}{\cal P}_{B\bar{B}}^{i}),
\end{equation}
where $N_c$ is the total number of input events  in category $c$. In the last term of this formula we have assumed $f_{{\rm sig},c} = f_{B\bar{B},c}$. 
The total likelihood function for all categories is given as the
product over the seven tagging categories (including a category for
 untagged events for yield determinations).
The product is extended to additional sub-decays by multiplying the above
product likelihoods for each decay.

We maximize the likelihood function while varying a set of free parameters: $S$, $C$, 
three background \xf\ PDF parameters, and, for each sub-decay,
signal and background yields, \DE\  and \mes\ background parameters, and
six parameters representing the background \deltat\ shape.

\section{Results}
\label{sec:Physics}
The results of  the fits to the neutral modes are shown in Table~
\ref{tab:resultsN}. We combine the five sub-decay modes shown in 
Table~\ref{tab:resultsN}. The decay mode 
$B^0 \rightarrow \etapr_{\eta_{3\pi}\pi\pi} K^0 (\pi^{0} \pi^{0})$ has 
not been reconstructed because of its expected low signal yield.
  The results for the three charged modes, used as a crosscheck, 
are presented in Table~\ref{tab:resultsC}; the values of $S$ and $C$ 
are consistent with zero, as expected.
The first two columns in each table represent the primary modes that we 
reported on in our previous analysis.  The last three columns of Table 
\ref{tab:resultsN} and the last column of Table \ref{tab:resultsC} are
for decay modes that we have not reported on previously.
The efficiency of the selection after all cuts is about 25\% for the
four primary decay modes and 15\% for the four new decay channels.

\begin{table}[htbp]
\caption{\hspace*{-0.2cm} Results from fits to  sub-decay modes of \BetapKzs. }
\label{tab:resultsN}
\begin{center}
\begin{tabular}{|l|c|c|c|c|c|}
\hline\hline
&$\etapr_{\eta\pi\pi} K^0 (\pi^{+} \pi^{-})$&$\etapr_{\rho\gamma} K^0 (\pi^{+} \pi^{-})$&$\etapr_{\eta_{3\pi}\pi\pi} K^0 (\pi^{+} \pi^{-})$ &$\etapr_{\eta\pi\pi} K^0 (\pi^{0} \pi^{0})$&$\etapr_{\rho\gamma} K^0 (\pi^{0} \pi^{0})$\\
\hline
Signal yield        & $192\pm15$ & $438\pm27$ & $55\pm9$ &$ 50\pm9$                  &$86\pm20$  \\ 
\BB\ yield          & $-$            & $93\pm33$     & $-$          &$-$             &$78\pm23$  \\ 
$S$                 & $0.05\pm0.28$  & $0.41\pm0.19$  & $0.53\pm0.49$&$-0.18\pm0.50$  &$-0.26\pm0.61$  \\ 
$C$                 & $-0.12\pm0.18$  & $-0.29\pm0.13$  & $0.18\pm0.38$&$-0.69\pm0.40$&$0.19\pm0.44$  \\ 
\hline
 \multicolumn{6}{|c|}{ {\bf  Combined  fit}} \\
\hline
Signal Yield        & \multicolumn{5}{c|}{$819\pm 38$} \\
$S$                 & \multicolumn{5}{c|}{$0.27\pm0.14$} \\
$C$                 & \multicolumn{5}{c|}{$-0.21\pm0.10$} \\
\hline\hline
\end{tabular}
\end{center}
\vspace*{-0.5cm}
\end{table}

\begin{table}[htbp]
\caption{\hspace*{-0.2cm} Results of fits to sub-decay modes \BetapK.}
\label{tab:resultsC}
\begin{center}
\begin{tabular}{|l|c|c|c|}
\hline\hline
&$\etapr_{\eta\pi\pi} K^+$ &$\etapr_{\rho\gamma} K^+$ &$\etapr_{\eta_{(3\pi)}\pi
\pi} K^+$ \\
\hline
Signal yield        & $585\pm26$       & $1322\pm48$     & $221\pm17$ \\
\BB\ yield          & $-$              & $774\pm75$      & $-$          \\
$S$                 & $-0.13\pm0.13$   & $-0.05\pm0.10$  & $-0.31\pm0.23$\\
$C$                 & $0.00\pm0.10$    & $-0.10\pm0.08$  & $-0.06\pm0.16$\\
\hline
 \multicolumn{4}{|c|}{ {\bf  Combined   fit}} \\
\hline
Signal Yield        & \multicolumn{3}{c|}{$2124\pm57$} \\
$S$                 & \multicolumn{3}{c|}{$-0.10\pm0.07$} \\
$C$                 & \multicolumn{3}{c|}{$-0.05\pm0.06$} \\
\hline\hline
\end{tabular}
\end{center}
\vspace*{-0.5cm}
\end{table}

\begin{figure}[!htbp]
\vspace*{-0.2cm}
\begin{center}
\includegraphics[angle=0,scale=0.50]{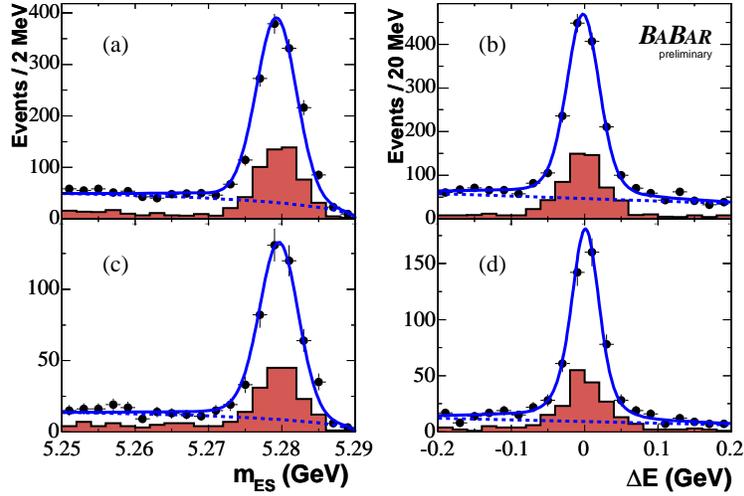}
\vspace*{-0.5cm}
\caption{ 
The $B$ candidate \mes\ and \DE\ projections for \BetapK\ (a, b) and
\BetapKz (c, d) in main decay modes . Points with errors represent the data, solid curves the full fit functions,
and dashed curves the background functions; the shaded histogram
represents the  $\eta^\prime_{\eta\pi\pi} K $ subset. }
\label{fig:projMbDE}
\end{center}
\end{figure}

\begin{figure}[!htbp]
\vspace*{-0.4cm}
\begin{center}
\includegraphics[bb=0 32 583 622,angle=270,scale=0.38]{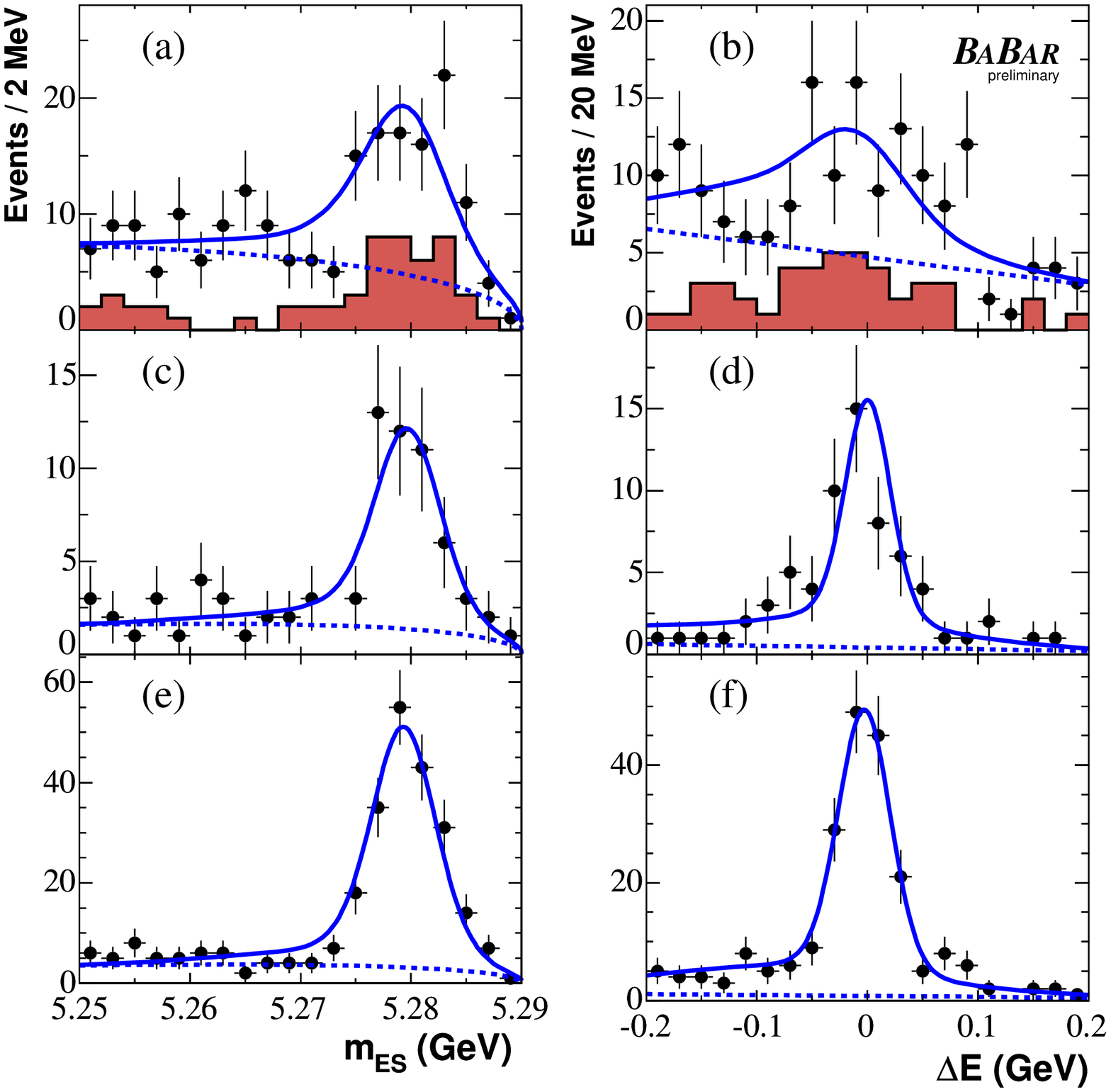}
\vspace{-0.2cm}
\caption{
$B$ candidate \mes\ and \DE\ projections for $\etapKz (\to\piz\piz)$ (a, b),
$\Bz\to\etapr_{\eta(3\pi)\pi\pi}\kzs (\to \pipi)$ (c, d), and
$\Bp\to\etapr_{\eta(3\pi)\pi\pi}\Kp$ (e, f).
Points with errors represent the data, solid curves the full fit functions,
and dashed curves the background functions. The shaded histogram in (a, b)
represents the  $\etapr_{\eta(\gaga)\pi\pi}\kzs\ $ subset. }
\label{fig:projMbDENew}
\end{center}
\end{figure}

In Fig.~\ref{fig:projMbDE}\ we show projections onto \mes\ and \DE\ of a
subset of the data of the primary decay modes for which the signal likelihood
(computed without the plotted variable) exceeds a mode-dependent
threshold that optimizes the sensitivity.
In Fig.\ \ref{fig:projMbDENew}\ we show the same  projections onto \mes\ and 
\DE\ for the new decay modes considered in the present  analysis.

\begin{figure}[!htbp]
  \begin{center}
  \includegraphics[bb=12 39 475 774 ,scale=0.70]{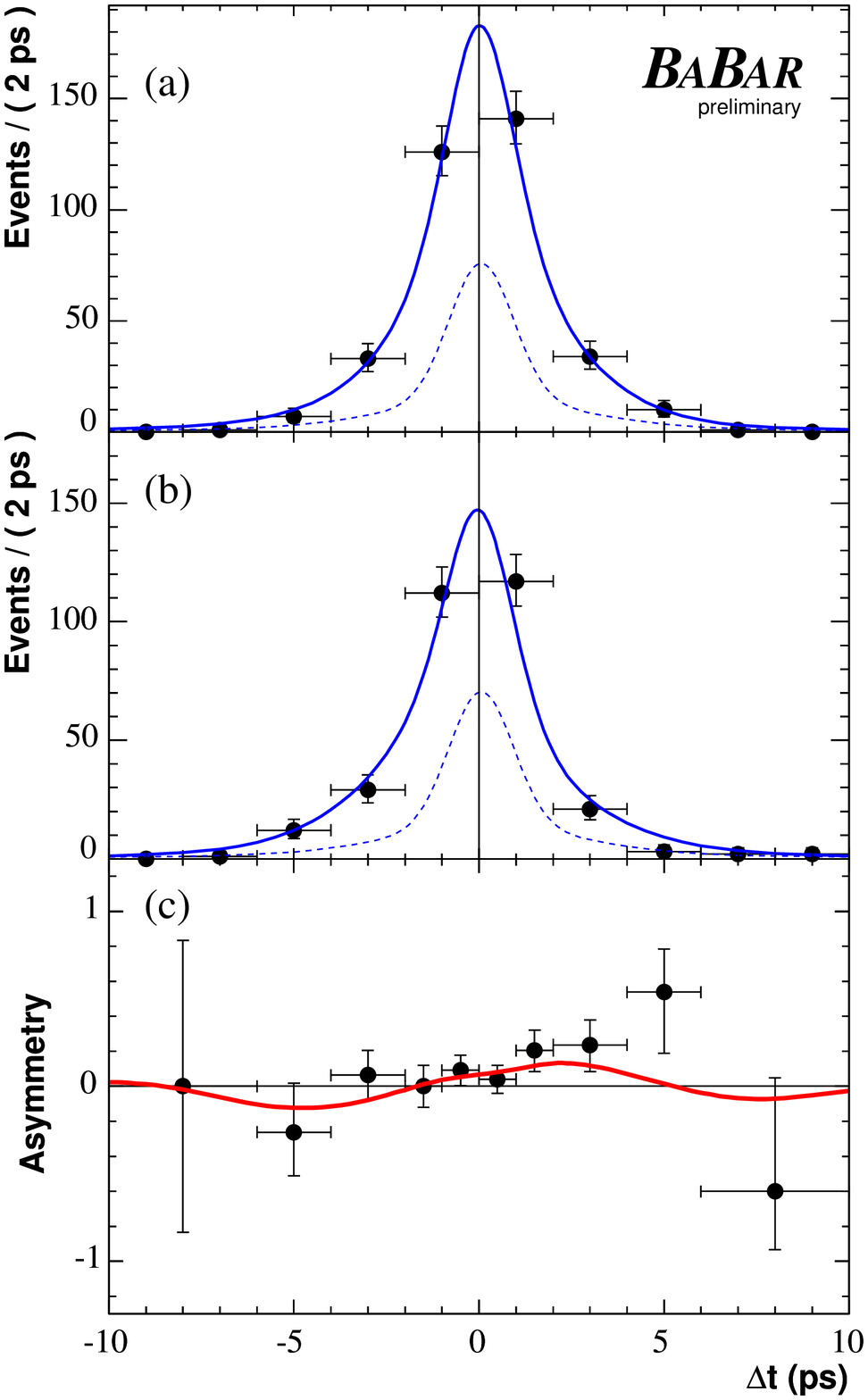}
\end{center}
 \caption{Projections onto \deltat\ for \BetapKzs, showing the  data (points with errors), fit function (solid line), and background function (dashed line), for (a) \Bz\ and (b) \Bzb\ tagged events and (c) the asymmetry between \Bz\ and \Bzb\ tags.}
  \label{fig:DeltaTProj}
\end{figure}

\begin{figure}[!htbp]
 \begin{minipage}{\linewidth}
  \begin{center}
  \includegraphics[bb=85 155 535 605 ,angle=-90,scale=0.35]{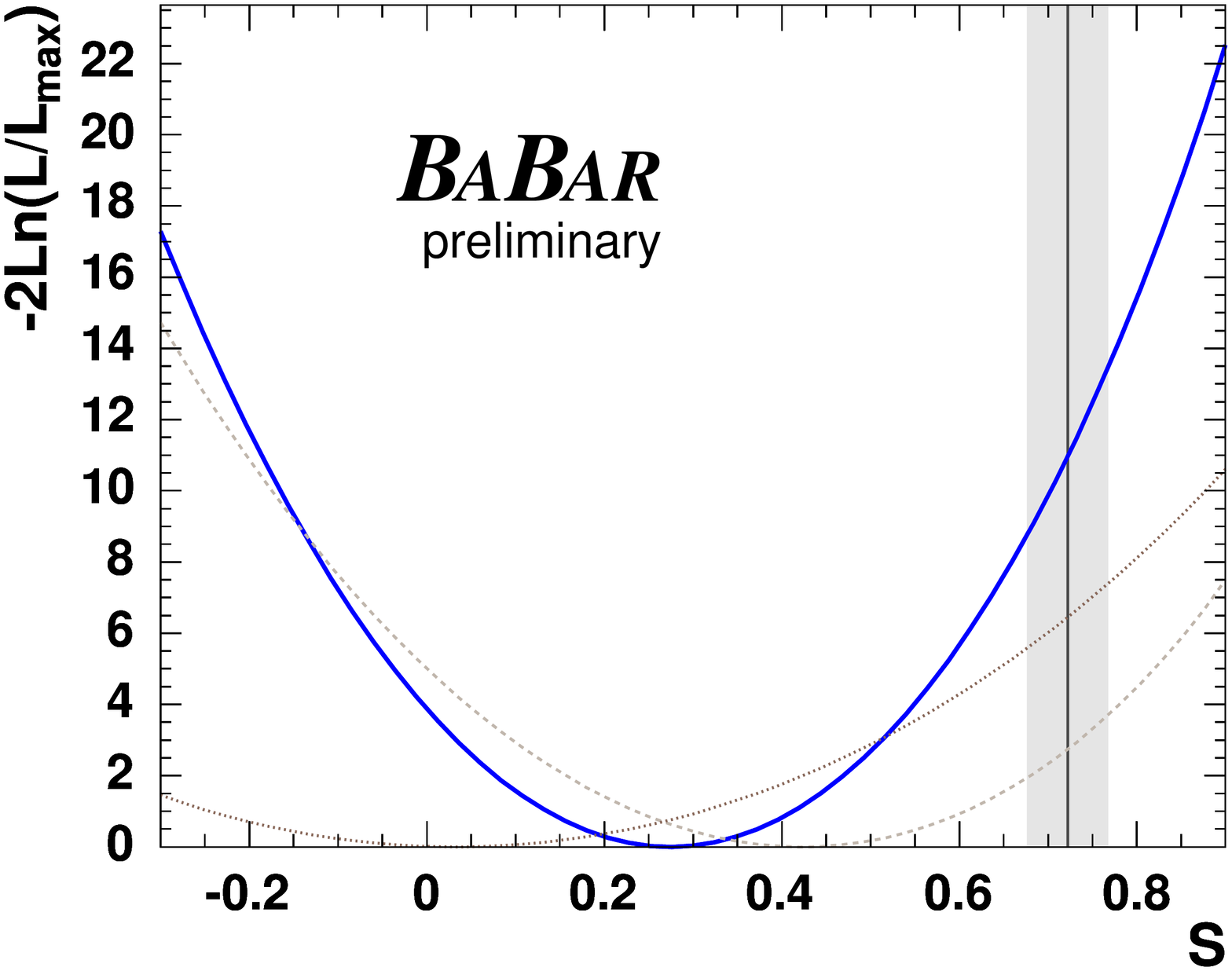}
  \hspace{3cm}
  \includegraphics[bb=85 155 535 605 ,angle=-90,scale=0.35]{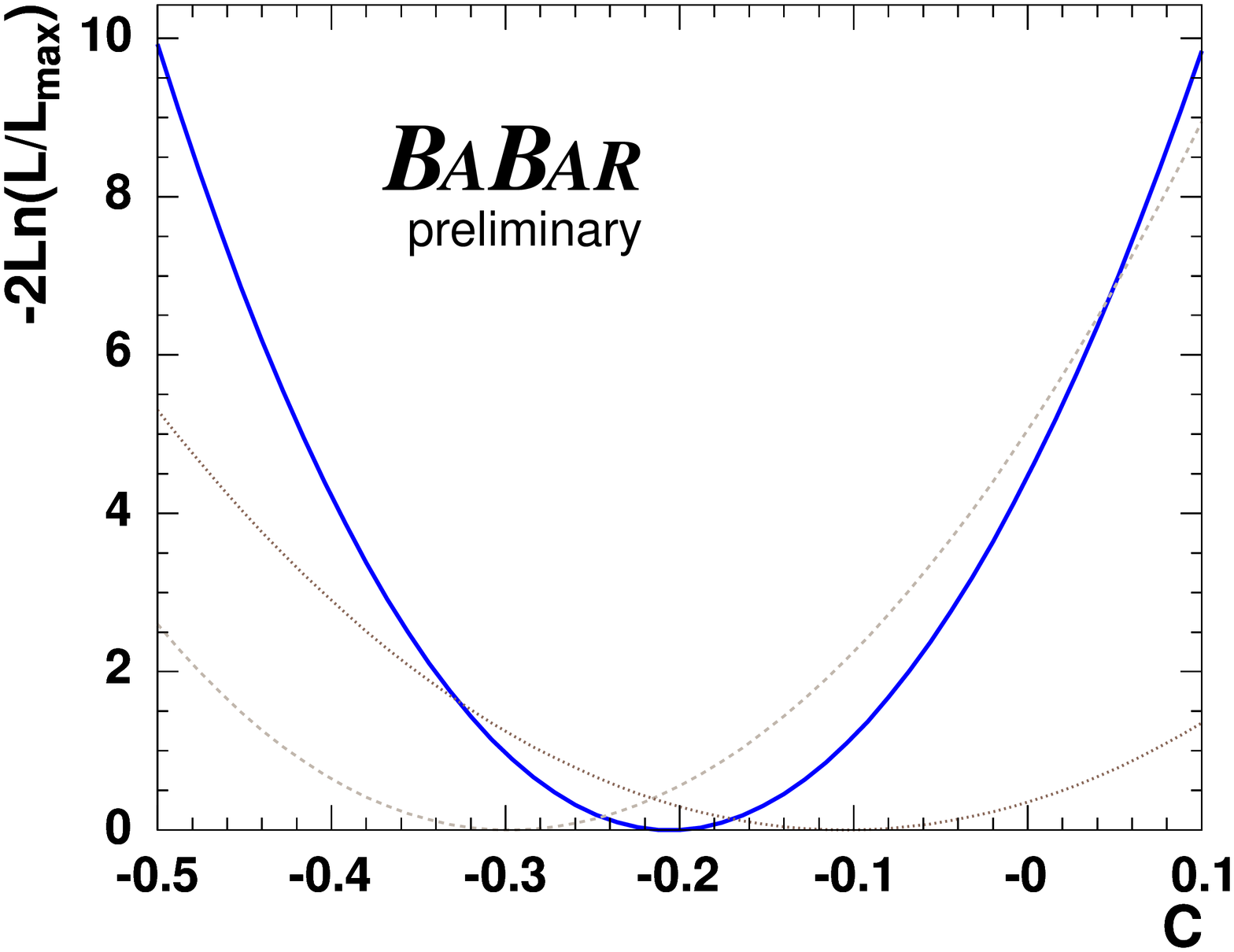}
   \end{center}
  \end{minipage}
  \vspace{0.2cm}
\caption{$-2\ln \calL$ scan for $S$ (left) and $C$ (right) parameters.
The solid blue line refers to all combined neutral sub-decays, the dotted line to the
$B^0 \rightarrow \eta^{\prime}_{\eta \pi \pi} K^0_S$ sub-decay
and the dashed line to the
$B^0 \rightarrow \eta^{\prime}_{\rho \gamma} K^0_S$ sub-decay.
The shaded band shows the \babar\ value of $\sin2\beta$ in charmonium decays \cite{s2b}.
}
  \label{fig:NLL}
\end{figure}

We show in Fig.~\ref{fig:DeltaTProj} the $\Delta t$
projections and asymmetry of  all  combined neutral modes for events
selected as for Fig.~\ref{fig:projMbDE} and Fig.~\ref{fig:projMbDENew}.
In Fig.~\ref{fig:NLL} we show the $-2\ln \calL$ scans for $S$ and $C$.
The best fit value for $S$ is 3.0 standard deviations from the \babar\
value of $\sin2\beta$ in charmonium decays \cite{s2b}.

\section{Systematic Uncertainties and Crosschecks}
\label{sec:Systematics}

The contributions to the systematic uncertainties in $S$ and $C$ are summarized in 
Table~\ref{tab:systtab}.
We evaluate the uncertainties associated with the PDF shapes by variation of
the parameters describing each discriminating variable. Systematic errors 
associated with signal parameters 
(\deltat\ resolution function, tagging fractions, and dilutions) are determined 
by varying their values within their errors.
Uncertainties due to \deltamd\ and $\tau_B$ are obtained 
by varying these parameters by the uncertainty in their
world average values \cite{PDG2004}. All changes are combined in quadrature
obtaining an error of 0.01 for both $S$ and $C$.

We vary the SVT alignment parameters in the signal Monte Carlo events by the size of misalignments found in the real data, and assign the resulting shift in the fit results as the systematic error.

The systematic errors due to interference between the 
CKM-suppressed $\bar{b} \rightarrow \bar{u} c \bar{d}$ amplitude and the 
favored $b \rightarrow  c \bar{u}d$
for some tag-side $B$ decays is found to be negligible for $S$ and gives a 
contribution to the $C$ uncertainty of about 0.012. The effect of \BB\ 
background is estimated to be negligible, but we assign 
an uncertainty of 0.01 in $S$ due to statistical limitations of this statement.
An uncertainty of 0.02 is assigned to account for limitations of Monte Carlo 
statistics and modeling of the signal.  We assign an uncertainty of 0.01
to account for the uncertainty in the position and size of the beam
spot, determined from variation of these quantities in signal MC.  The total systematic error is obtained by summing individual errors in quadrature.

\begin{table}[htbp]
\caption{Estimates of systematic errors.}
\label{tab:systtab}
\begin{center}
\begin{tabular}{lcc}
\hline\hline
Source of error &  $\sigma(S)$ & $\sigma(C)$  \\
\hline
PDF Shapes              &$0.01$  &$0.01$     \\
SVT alignment           &$0.01$  &$0.01$    \\
Tag-side interference   &$0.00$  &$0.01$ \\
\BB\ Background         &$0.01$  &$0.00$    \\
MC statistics/modeling  &$0.02$  &$0.02$    \\
Beam spot               &$0.01$  &$0.01$    \\
\hline
Total                   &$0.03$  &$0.03$    \\
\hline\hline
\end{tabular}
\end{center}
\end{table}

We have also performed a number of checks of our results.  We divide the
sample into two sub-samples --- the previously published sample and the data
collected in 2003/2004.  We find consistency with our previous results
and between the two sub-samples.  When we fit with the value for $C$
fixed to zero, we find $S= 0.29\pm0.14$.  We produce samples of
pseudo-experiments generated with events produced to match the PDF
distributions.  From these samples , we verify that the  fit bias on $S$ and $C$  is negligible and that there is a good agreement between expected and observed errors.

\section{Conclusion}
We have reconstructed  $2124 \pm 57$ \BetapK\ events and $819\pm38$ 
\BetapKzs\ events, about two-thirds of which have a flavor tag.  We have 
used the latter sample
to measure the time-dependent CP-violating parameters for \BetapKzs. We find 
$S= 0.27 \pm 0.14 ({\rm stat}) \pm 0.03 ({\rm syst})$  and 
$C= -0.21 \pm 0.10 ({\rm stat}) \pm 0.03 ({\rm syst})$. 

The measured value of $S$ is 3.0 standard deviations from the the \babar\ 
measurement of \stwob, $0.722\pm0.040\pm0.023$, from $B\to {\rm charmonium}$ \KS
decays \cite{s2b}.  The observed deviation $\Delta S$ is large compared with 
the expected theoretical uncertainty.

\section{Acknowledgments}
\label{sec:Acknowledgments}


We are grateful for the 
extraordinary contributions of our \pep2\ colleagues in
achieving the excellent luminosity and machine conditions
that have made this work possible.
The success of this project also relies critically on the 
expertise and dedication of the computing organizations that 
support \babar.
The collaborating institutions wish to thank 
SLAC for its support and the kind hospitality extended to them. 
This work is supported by the
US Department of Energy
and National Science Foundation, the
Natural Sciences and Engineering Research Council (Canada),
Institute of High Energy Physics (China), the
Commissariat \`a l'Energie Atomique and
Institut National de Physique Nucl\'eaire et de Physique des Particules
(France), the
Bundesministerium f\"ur Bildung und Forschung and
Deutsche Forschungsgemeinschaft
(Germany), the
Istituto Nazionale di Fisica Nucleare (Italy),
the Foundation for Fundamental Research on Matter (The Netherlands),
the Research Council of Norway, the
Ministry of Science and Technology of the Russian Federation, and the
Particle Physics and Astronomy Research Council (United Kingdom). 
Individuals have received support from 
CONACyT (Mexico),
the A. P. Sloan Foundation, 
the Research Corporation,
and the Alexander von Humboldt Foundation.

\end{document}